\documentstyle[aps]{revtex}
\input{psfig}
\twocolumn

\begin{document}
\title{ Universal property of the information entropy in atoms, nuclei
and atomic clusters}

\author{S.E. Massen and  C.P. Panos}

\address{Department of Theoretical Physics
Aristotle University of Thessaloniki,
GR 54006 Thessaloniki, Greece}

\maketitle

\begin{abstract}
The position-- and momentum--space information entropies of the
electron distributions of atomic clusters are calculated using a
Woods--Saxon single particle potential. The same entropies are
also calculated for nuclear distributions according to the
Skyrme parametrization of the nuclear mean field. It turns out
that a similar functional form $S=a+b \ln N$ for the entropy as
function of the number of particles $N$ holds approximately for
atoms, nuclei and atomic clusters. It is conjectured that this
is a universal property of a many-fermion system in a mean
field. It is also seen that there is an analogy of
our expression for $S$ to Boltzmann's thermodynamic entropy
$S=k \ln W$.

\end{abstract}

\vspace{3mm}

PACs: 89.70.+c;  36.40.+d;  31.10.+z;   21.60.-n

\vspace{3mm}

Information--theoretical methods have played in recent years an important
role in the study of quantum mechanical systems
\cite{Bialynicki75,Gadre84,Gadre85,Gadre87,Ohya93,Nagy96,Majernic96,Panos97,Lalazis98} in two cases: first in
the clarification of fundamental concepts of quantum mechanics
and second in the synthesis of probability densities in position
and momentum spaces. In the first case an important step was the
discovery of an entropic uncertainty relation (EUR)
by Bialynicki--Birula and Mycielski  \cite{Bialynicki75}
which for a three--dimensional system has the form:
\begin{equation}
S_r+S_k \ge 3 (1+\ln \pi)\, , \quad (\hbar=1)
\label{(1)}
\end{equation}
(see also ref. \cite{Hirsc57} for the one--dimensional case). In
(\ref{(1)})
 $S_r$ is the Shannon information entropy in position--space:
\begin{equation}
S_r=-\int \rho(\vec{ r}) \ln \rho( \vec{ r}) \ d \vec{r}
\label{(2)}
\end{equation}
$S_k$ is the corresponding entropy in momentum-space:
\begin{equation}
S_k=-\int n(\vec{k}) \ln n(\vec{k}) \ d\vec{k}
\label{(3)}
\end{equation}
and $\rho(\vec{r})$, $n(\vec{k})$ are the position-- and momentum--space
density distributions respectively, which are normalized to one.
However, for a normalization to the number of particles $N$,
the following EUR holds  \cite{Gadre87}:
\begin{equation}
S_r+S_k \ge 3N(1+\ln \pi)-2N \ln N=6.434 N - 2 N \ln N
\label{(4)}
\end{equation}

Inequality (1), for the information entropy sum in conjugate
spaces, is a joint measure of uncertainty of a quantum mechanical
distribution, since a highly localized $\rho(\vec{r})$ is
associated with a diffuse $n(\vec{k})$, leading to low $S_r$ and
high $S_k$ and vice--versa. Expression (\ref{(1)}) is an
information--theoretical uncertainty relation stronger  than
Heisenberg's \cite{Bialynicki75}.
We also note that expression (\ref{(1)}) does not depend
on the unit of length in measuring $\rho(\vec{r})$ and $n(\vec{k})$ i.e.
the sum $S_r+S_k$ is invariant to uniform scaling of coordinates.

 Gadre \cite{Gadre84} derived the following approximate expression for
the information entropies of electron distributions in atoms:
\begin{equation}
S_r+S_k \simeq 6.65 N - N \ln N
\label{(5)}
\end{equation}
using Thomas-Fermi theory
and Gadre et al \cite{Gadre85} derived :
\begin{equation}
S_r+S_k \simeq 6.257 N - 0.993 N \ln N
\label{(6)}
\end{equation}
with Hartee-Fock calculations. Here, $N$ is the number of electrons.

Panos and Massen \cite{Panos97} found the following expression for nuclear
distributions, employing the simple harmonic oscillator (HO)
model of the nucleus:
\begin{equation}
S_r+S_k \simeq 5.287N-1.13N \ln N
\label{(7)}
\end{equation}
where $N$ is the number of nucleons in nuclei. Relations of the same
functional form hold for $S_r$ and $S_k$ separately but the important
quantity is $S_r + S_k$.

There is a striking similarity of (5), (6) and (7) with the EUR (4),
indicating that the functional form
\begin{equation}
S = a N + b N \ln N
\label{(7a)}
\end{equation}
is universal for a many-fermion system in a mean field.

However, the above relations were derived for a normalization of
$\rho(\vec{r})$ and $n(\vec{k})$ to the number of particles $N$.
In the following we
find it more convenient to normalize to one . There is a simple
relationship between the two cases and we can easily transform
one case to the other according to the relations:
$$S_r[norm=1]=\frac{S_r[norm=N]}{N}+\ln N$$
$$
\quad S_k[norm=1]=\frac{S_k[norm=N]}{N}+\ln N
$$
Hence, we have for normalization to one,
the following expressions:
\begin{equation}
S_r+S_k \simeq 6.65+ \ln N \quad (atoms, Thomas-Fermi)
\label{(8)}
\end{equation}

\begin{equation}
S_r+S_k \simeq 6.257+1.007 \ln N \quad (atoms, Hartee-Fock)
\label{(9)}
\end{equation}

\begin{equation}
S_r+S_k \simeq 5.287+0.870 \ln N  \quad (nuclei-H.O.)
\label{(10)}
\end{equation}

In the present letter we extend our calculations for two other cases:
the  distribution of the valence electrons in atomic clusters using a
Woods--Saxon
single particle potential and the nuclear distribution in nuclei
employing the Skyrme parametrization of the nuclear mean field.

In atomic (metallic) clusters the effective radial electronic potential was
derived by Ekardt \cite{Ekardt84} in his
spherical--jellium--background--model study of the self--consistent
charge density and the self--consistent effective one--particle potential,
using the local density approximations.
Ekardt's potentials for neutral sodium clusters
were parametrized in ref. \cite{Nishioka90} by a
Woods--Saxon potential of the form:

\begin{equation}
V_{WS}(r)= - \frac{V_0}{1 + \exp [ \frac{r-R}{a}]}
\label{(11)}
\end{equation}
with $V_0=6 \ eV$, $R=r_0 N^{1/3}$, $r_0= 2.25 \ \AA$ and $a=0.74\ \AA$.
For a detailed study regarding the parametrization of Ekardt's potentials see
ref. \cite{Kotsos97}.

We solved numerically the Schr\"{o}dinger equation
for atomic
clusters with $Z=8$, $18$, $20$, $34$, $40$, $58$, $68$, $70$ valence 
electrons in the potential (\ref{(11)}) and found the wave functions of the
single--particle states in configuration space and by Fourier transform
the corresponding ones in momentum space. Using the above wave functions,
we calculated the electron density $\rho(\vec{r})$ in position space and
$n(\vec{k})$ in momentum space, which were inserted into equations
(\ref{(2)}), (\ref{(3)})  and gave us the values of the information
entropies $S_r$ and $S_k$ .
Then we fitted the form $S=a+b \ln N$ to these values  and
obtained the expressions:
\begin{eqnarray}
S_r \simeq     4.133 + 0.934  \ln N        \label{(12)} \\
S_k \simeq     1.563 - 0.027  \ln N        \label{(13)}    \\
S_r+S_k \simeq 5.695 + 0.907  \ln N        \label{(14)}
\end{eqnarray}

Next the nuclear densities $\rho(\vec{r})$ and $n(\vec{k})$
for several nuclei were obtained with Hartee--Fock calculations
using the Skyrme parametrization of the nuclear mean
field. There are various parametrizations of the Skyrme interaction,
but they affect slightly the information entropies \cite{Lalazis98}.
Thus we used the SKIII interaction \cite{Skyrme}. Finally, we fitted
the form $S=a+b\ln N$ to the values obtained with SKIII interaction and
we found the expressions:
\begin{eqnarray}
S_r& \simeq &3.395+0.767 \ln N        \label{(15)}  \\
S_k& \simeq &1.929+0.091 \ln N          \label{(16)}   \\
S_r + S_k& \simeq & 5.319+0.860 \ln N   \label{(17)}
\end{eqnarray}

The fit is in reasonably good agreement with its H.O. counterpart
(comparison of relation (\ref{(17)}) to (\ref{(10)})), though the
individual entropies $S_r$ and $S_k$ do not match with the respective
HO ones \cite{Panos97} that well. It seems that there is a delicate
balance between the coordinate and momentum spaces, so that the
interesting quantity is the sum $S=S_r+S_k$ (the net information
content of the system) and not the individual entropies $S_r$ and $S_k$.
\begin{figure}
\centerline{\psfig{figure=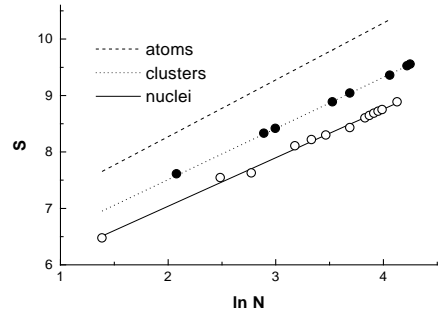,width=6.5cm} }
\caption{Information entropy $S$ as function of the number of particles $N$
according to our fitted expression $S=a+b \ln N$ for atoms,
clusters and nuclei. The corresponding values
of our numerical calculations are denoted by solid circles (clusters) and
open circles (nuclei).}
\end{figure}

In figure 1 we plot our fitted form $S=S_r+S_k=a+b \ln N$ for atoms (with
Hartee-Fock, relation (\ref{(9)}), upper curve), atomic neutral Na
clusters (relation
(\ref{(14)}), middle curve) and nuclei (with Skyrme, relation (\ref{(17)}),
lower curve). These lines correspond to our fitted expressions, while the
corresponding values of our numerical calculations are denoted
by solid circles (clusters) and open circles (nuclei with SKIII
interaction).

Concluding, in the present letter we derive an interesting characteristic of
information entropies $S_r$ and $S_k$ for various systems i.e. atoms
(Thomas--Fermi theory, Hartee--Fock), nuclei (Harmonic Oscillator
model, Skyrme force) and atomic clusters(Woods--Saxon potential).
For all of these systems the entropies can be represented well
by a function, which incorporates $\ln N$ linearly i.e.
$S= a + b \ln N $
where $N$ is the number of electrons in atoms or
nucleons in nuclei or electrons in atomic clusters. We may
conjecture that this is a universal property of a many-fermion
system in a mean field. As stated in ref. \cite{Gadre85}, the information
entropies seem to be a hidden treasure, as yet remains mostly
unexplored.

A final comment seems appropriate: There is an analogy
of our expression $S= a+b \ln N$
to Boltzmann's thermodynamic entropy
$S=k \ln W$ ($k$ is Boltzmann's constant).
This can be seen as follows:
Our relation can be written in the form:
\begin{equation}
S=b \ln (N/ N_0)
\label{(s3)}
\end{equation}
where $1/N_0= e^{a/b}$.
The above entropy  is the information entropy $S_{inf}$, which
is related to the physical entropy $S_{phys}$ through Jaynes'
relation: $S_{phys}= k S_{inf}$. In our case, we have:
\begin{equation}
S_{phys} = k b \ln (N/N_0)
\label{(s4)}
\end{equation}
Here $b$ and $N_0$ depend on the system under consideration (atom,
cluster or nucleus), but this dependence is not strong.
One can take that $b \approx 1$ (actually equals $1.007$ for atoms,
0.907 for clusters and 0.860 for nuclei) and $1/N_0 \approx 500$ (because
$1/N_0=500$ for atoms, $485$ for nuclei and $533$ for clusters).
Thus we obtain the following rough estimate for the entropy
of a many--fermion system:
\begin{equation}
S_{phys}  \approx  k \ln (500 N)
\label{(18)}
\end{equation}
The number $500 N$ corresponds to the number of equiprobable
microstates $W$ of Boltzmann's relation.

Hence, starting from Shannon's information entropies of
quantum--mechanical distributions, we arrived at relation
(\ref{(18)}), which can be considered as a quantum-mechanical
analogue to Boltzmann's entropy $S=k \ln W$.

\section*{Acknowledgments}

The authors would like to thank Professor M.E. Grypeos for an informative
discussion in atomic clusters


\end{document}